%% file: lochin-kred.tex
\documentclass[conference]{IEEEtran}
\usepackage{epsfig}
\usepackage{graphics}
\usepackage{url}
\usepackage{color}
\usepackage{subfigure}
\def\imgsz {0.40\columnwidth}
\def\imgszfig {0.90\columnwidth}

\newcommand{\up}[1]{\textsuperscript{#1}}

\title{Managing network congestion with a Kohonen-based RED queue}

\author{\authorblockN{Emmanuel Lochin}
\authorblockA{
Universit\'{e} de Toulouse, DMIA\\
ISAE - France,\\
emmanuel.lochin@isae.fr}
\and
\authorblockN{Bruno Talavera}
\authorblockA{Universit\'{e} Pierre et Marie Curie,\\
Polytech'Paris, France,\\
bruno.talavera@upmc.fr}}

\date{}
\begin{document}
\maketitle
\thispagestyle{empty}

\begin{abstract}
The behaviour of the TCP AIMD algorithm is known to cause queue length oscillations when congestion occurs at a router output link. 
Indeed, due to these queueing variations, end-to-end applications experience large delay jitter. Many studies have proposed efficient Active 
Queue Management (AQM) mechanisms in order to reduce queue oscillations and stabilize the queue length. These AQM are mostly 
improvements of the Random Early Detection (RED) model. Unfortunately, these enhancements do not react in a similar manner 
for various network conditions and are strongly sensitive to their initial setting parameters.
Although this paper proposes a solution to overcome the difficulties of setting these parameters by using a Kohonen neural
network model, 
another goal of this study is to investigate whether cognitive intelligence could be placed in the core network to solve such stability problem.
In our context, we use results from the neural network area to demonstrate that our proposal, named Kohonen-RED (KRED), 
enables a stable queue length without complex parameters setting and passive measurements.
\end{abstract}

%-------------------------------------------------------------

\section{Introduction}

More than ten years ago, the Random Early Detection (RED) was proposed to avoid congested links \cite{floyd93red}.
The main idea of the RED algorithm is to drop packets before the queue is full. As a consequence, when a TCP source gets such preventive 
drops, it decreases the emitted throughput according to the AIMD (Additive Increase Multiplicative Decrease) 
algorithm. RED drops packets with an increasing probability ($max_{p}$) when the occupancy of the queue lies between two thresholds ($min_{th}, max_{th}$). 
The goal of RED is to maintain a small buffer occupancy and avoid casual bursts of packet losses. 

The authors in \cite{may99reasons} and \cite{ziegler01red} weight up the disadvantages for deploying such mechanism. 
In certain cases, increasing the number of dropped packets can have unexpected effects on the overall performance \cite{ziegler01red}. 
This has motivated the use of preventive marking instead of preventive dropping with the use 
of the Efficient Congestion Notification (ECN) flag. In this case, instead of dropping packets, the RED queue marks the packet's ECN flag 
to notify senders that they are crossing a congested link and that they should decrease their sending rate.
In \cite{may99reasons}, the authors claim that tuning parameters in RED remains an inexact science.
We fully acknowledge the criticisms of this approach which motivate our proposal of managing the RED configuration with a neural network. 

Feng RED (FRED) \cite{feng99self} and Adaptive RED (ARED) \cite{floyd01adaptive} introduced the notion of adaptive AQM. These adaptive strategies 
recompute the $max_{p}$ probability value following an AIMD algorithm.
However, the parameters that weight this AIMD process remain difficult to estimate. 

Some past work have already suggested that RED is fundamentally hard to tune \cite{low03linear}. In this study, the authors show that RED 
parameters can be tuned to improve stability, but only at the cost of large queues even when they are dynamically adjusted.
Even if other different queueing approaches have been proposed to improve the efficiency of RED-like algorithms in various 
network conditions, the parameters used to set these new AQM are sometimes more complex to determine than RED. 
In particular, this is the case for the PI controller \cite{hollot01designing}. Nowadays, general parameters able to stabilize the queue don't 
yet exist whatever the AQM used and we could discuss whether the problem is in fact solvable.

Although the validity of RED concept is still debated, we claim that the parameters' settings are one of the main barrier to its acceptance.
In this paper, we propose to compute the optimal $max_{p}$ value with a Kohonen neural network \cite{kohonen}.
We do not attempt to design another queueing mechanism or propose to enhance the core mechanism itself.
We only focus on the optimal estimation of the probability parameter. This paper aims at illustrating the impact of the role of learning mechanisms
on core network Internet problems with similar motivation than the one presented in \cite{csail07learning}.

This paper is structured as follow. Section \ref{sec:motiv} presents the motivation of this work.
Section \ref{sec:impl} gives pointers related to the implementation of the core mechanism. Section \ref{sec:train} presents
the training phase of the neural network. Then, section \ref{sec:eval} evaluates the proposal and
finally section \ref{sec:conclusion} gives the perspectives of this work.

\section{Motivation of using a Kohonen neural network}
\label{sec:motiv}

\begin{figure}[htb]
\begin{center}
\subfigure[Pole balancing]{
\includegraphics[ keepaspectratio=true, width=\imgszfig]{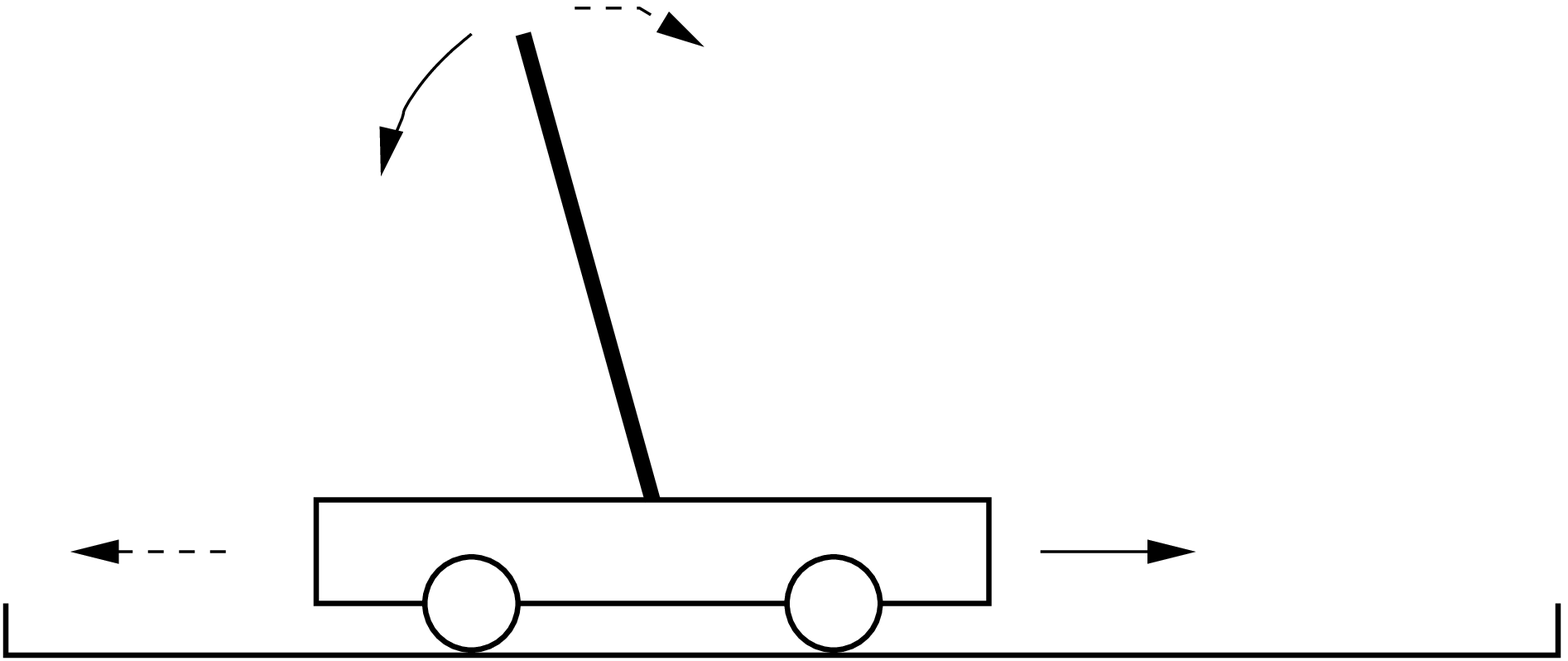}
\label{fig:pole}
}
~\\
~\\
\subfigure[Adaptive RED]{
\scalebox{0.45}{ 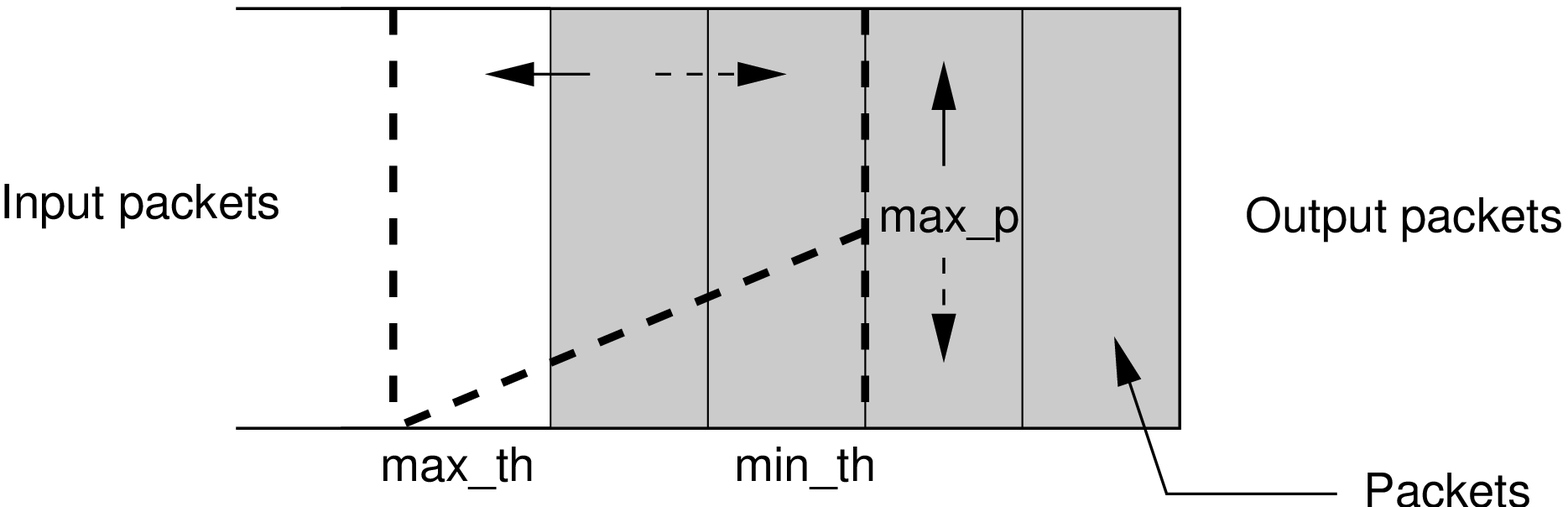 }
\label{fig:redpole}
}
\caption{Analogy between the single pole balancing problem and RED AQM}
\label{fig:analogy}
\end{center}
\end{figure}

Kohonen networks are a class of neural networks known to solve the pole balancing problem \cite{polebalancing}. 
Pole balancing is a control benchmark historically used 
in mechanical engineering. It involves a pole placed on a cart via a joint allowing movement along a single axis. 
The cart is able to move along a track with a fixed length as represented in figure \ref{fig:pole}.
The aim of the problem is to keep this pole balanced by applying forces to the cart.

The main idea of our contribution is based on the analogy existing between this balancing problem and the 
RED queueing problem. In RED, we can compare the pole balancing to the evolution of the queue occupancy which
oscillates between both thresholds ($min_{th}, max_{th}$). The physical forces resulting on the pole have a
similar role to the packets arrival rate in the queue. 
Figure \ref{fig:analogy} illustrates this view.

\begin{table}[ht!]
\begin{center}
\begin{tabular}{|l|c|c|}\hline
         & Pole & RED \\\hline
 & & \\
input\_value[1] & previous position & previous \\
                &                   & queue length\\
 & & \\
input\_value[2] & new position & current \\
                &              & queue length\\
 & & \\
output\_value[1]   & force to apply & $max_{p}$\\
  		   & in Newton & \\\hline
\end{tabular}
\end{center}
\caption{Input and output values used}
\label{tab:io}
\end{table}

Self-configuring RED schemes such as FRED or ARED update the $max_{p}$ value as a function of the
arrival rate in order to stabilize the queue size between both thresholds, $min_{th}$ and $max_{th}$. 
In \cite{feng99self}, the authors explain the queue length variation by the need of dynamically changing $max_{p}$ 
as function of the queue occupancy. They propose to recompute this probability following an AIMD algorithm.
The update is done as function of the average queue size. If the average queue size is around $max_{th}$, the algorithm 
increases $max_{p}$ to drop more packets and decreases $max_{p}$ if the value is around $min_{th}$. 

The AIMD algorithm performed by FRED is 
different from ARED. Indeed, FRED updates $max_{p}$ each time a packet is enqueued
while ARED has another parameter allowing to update this value during a time interval. This action period can smooth
the effect of an aggressive setting of the AIMD factors. Moreover, FRED does not apply consecutive decrease or consecutive increase
of the $max_{p}$ value. This choice can be problematic in case of rapid traffic change.

The neural network we use here is known as the Kohonen Self Organizing Map (SOM) \cite{kohonen}. It consists in a one or 
two dimensional information processing layer of functional entities called neurons. It is connected to input data seen as 
input vectors and provides output data also as vectors. We present in table \ref{tab:io} the entries used to feed the neural network
in both cases and the resulting output. The input vector contains the previous and the current queue length and the output vector the
$max_{p}$ probability. For a sake of comparison, we give in this table the vectors used with the pole balancing problem. 
The input data is fully mapped onto the Kohonen layer's neurons 
which respond to this data according to the weight assigned to the connexions between input vectors and neurons and deliver 
an output response vector. To begin with, the neural network is presented a learning set of example input vectors and adjusts 
(i.e. learns) appropriate weights for its neurons by comparing the input vectors to the weight vectors for each neuron thus 
electing a "winning" neuron "close" to the input vector. 

In addition to this, the Kohonen SOM deals with a topological learning feature, which implies neural neighborhood generalization 
of a correct learning experience so as to create clusters of neurons responding to similar input vectors without necessarily 
having explicitly learnt them. If a neuron learns that a given input vector is a vector it should respond to, its neighbours 
will learn they also should respond, only in a lesser way, depending on their topological distance to the first "winning" neuron. 
This way, the Kohonen SOM is well adapted to stability preservation tasks as the one we present here. Once the learning procedure 
is over, i.e. when the neural network produces an acceptable amount of erroneous responses during learning, the weights of the 
neural connexions to the data input are freezed. That means that the training process needs to be done only once without specific
scenario and should work for every kind of situation. 

Given the Kohonen SOM algorithm, the neural network can generalize its learnt 
experiences to other input vectors it has never seen before and produce adapted responses. In this way, the conservation of a 
direction, an equilibrium or the correct parameter to adjust a RED mechanism is made possible although there is no way of predicting 
the way the neural network learns to solve this particular problem. In our case, the learnt sequences of input vectors 
are not the ones used in our tests, in order to prove that the learning method provides a general purpose neural network for the 
resolution of the problem we deal with here. Once it has learnt, it can be used indefinitely for the task it has been trained for.

Previous related work \cite{cho05neural} presents the use of a multi layer perceptron to adapt the $\alpha$ and $\beta$ coefficients
of a PI controller. In \cite{cho05neural}, the authors don't improve the queue length stability but smooth the PI dynamic and in
average, results obtained are globally similar. 
We think that such a neural network is well adapted to pattern and shape 
recognition problems, whilst a SOM such as the Kohonen SOM could be better suited to the task of stability preservation which 
we deal with here. Indeed, this Kohonen SOM algorithm preserves topological relationships between neighbouring vectors.

Each time a packet is enqueued, the Kohonen network computes a new $max_{p}$ following the previous and the current 
average queue size. No other parameters are needed to perform this operation.

\section{Implementation}
\label{sec:impl}

One important point of dealing with Kohonen network is the small memory footprint required by the implementation. In our case, we have implemented our proposal in ns-2 simulator. The most complex structure is simply a square matrix $25 \times 25$ which represents the Kohonen network. The code used
is a modification of the well-known Karsten Kutza's implementation\footnote{http://www.neural-networks-at-your-fingertips.com/}.
All the scripts and ns-2 implementation used in this study are available for download at the following address: 
\url{http://mobqos.ee.unsw.edu.au/~lochin/kred.html}.

\section{The training process}
\label{sec:train}

The KRED queue has been trained with an arbitrary chosen number of eight long-lived TCP/Newreno flows emitted during 600 seconds without traffic variation 
on a single link topology. The neural network map learnt to stabilize 
the KRED queue with the common parameters given table \ref{tab:params} after 331 seconds. No further training has been done. The resulting
Kohonen map is used thereafter in all the experiments. 
Both experiments related in section \ref{sec:eval} use the same Kohonen map resulting from the same training process.

\section{Evaluation and analysis}
\label{sec:eval}

This section presents the experiments driven to evaluate the KRED and comments the results obtained.

\subsection{Testbed and assumptions}

\begin{figure}[hb!]
\begin{center}
\includegraphics[width=\columnwidth]{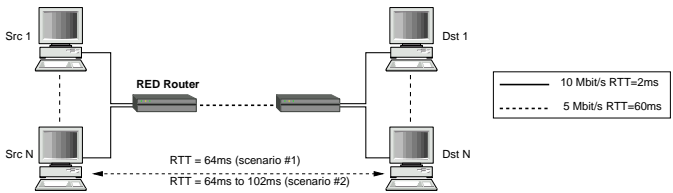}
\end{center}
\caption{The simulation topology}
\label{fig:testbed}
\end{figure}

\begin{figure}[hb!]
\begin{center}
\subfigure[First scenario]{\epsfig{figure=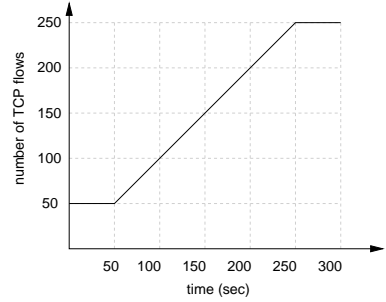, width=5cm}\label{fig:scenario1}}
\subfigure[Second scenario]{\epsfig{figure=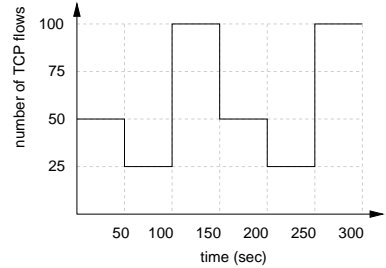, width=5cm}\label{fig:scenario2}}
\caption{The simulation scenarios}
\label{fig:scenarios}
\end{center}
\end{figure}

We drive experiments over a standard dumbbell topology represented in figure \ref{fig:testbed}. We compare our proposal to 
RED, Feng RED (FRED), ARED and PI AQM. The parameters used
for each queue are given in table \ref{tab:params}. The TCP flows are NewReno with a large window size set 
to $10000$ packets.  The RED queue is configured to drop and not to mark packets. 
In order to evaluate our proposal, we drive two distinct experiments. In the first scenario, the number of TCP flows
in the network is increasing from 50 to 250 flows following the pattern figure \ref{fig:scenario1}. The RTT for each flow is identical. This scenario allows us to verify the 
impact of the traffic load on our proposal compared to others AQM.
In the second experiment, the traffic changes every 50 seconds following the scenario presented in
figure \ref{fig:scenario2}. Furthermore, each flow has a random RTT ranging from 64 to 102 ms. The rationale for using this traffic pattern is to evaluate
our proposal under wide traffic variations.

\begin{table}[ht!]
\begin{center}
\begin{tabular}{c|l}
Common  & $min_{th}=100pkts$ $max_{th}=150pkts$,\\
Parameters (C.P.)		 & $q_{size}=200pkts$, $q_{weight}=10^{-4}$.\\
			 & \\
RED                      & C.P., $max_{p}=0.1$.\\
			 & \\
FRED                     & C.P., $max_{p}=0.1$ $\alpha = 3.0$, $\beta = 2.0$.\\
			 & \\
ARED                     & C.P.,  $\alpha = 0.01$, $\beta = 0.09$, $gentle=true$,\\
                         & $interval = 0.3$, $max_{p}=0.1$.\\
			 & \\
PI                       & $a=1.822.10^{-5}$, $b=1.816.10^{-5}$,\\
			 & $q_{ref}=100pkts$, $w=170Hz$.\\
			 & \\
KRED                     & C.P.
\end{tabular}
\end{center}
\caption{RED parameters used}
\label{tab:params}
\end{table}

\subsection{First scenario}

\begin{figure*}[ht!]
\begin{center}
\subfigure[RED]{
\includegraphics[ keepaspectratio=true, width=\imgsz, angle =-90]{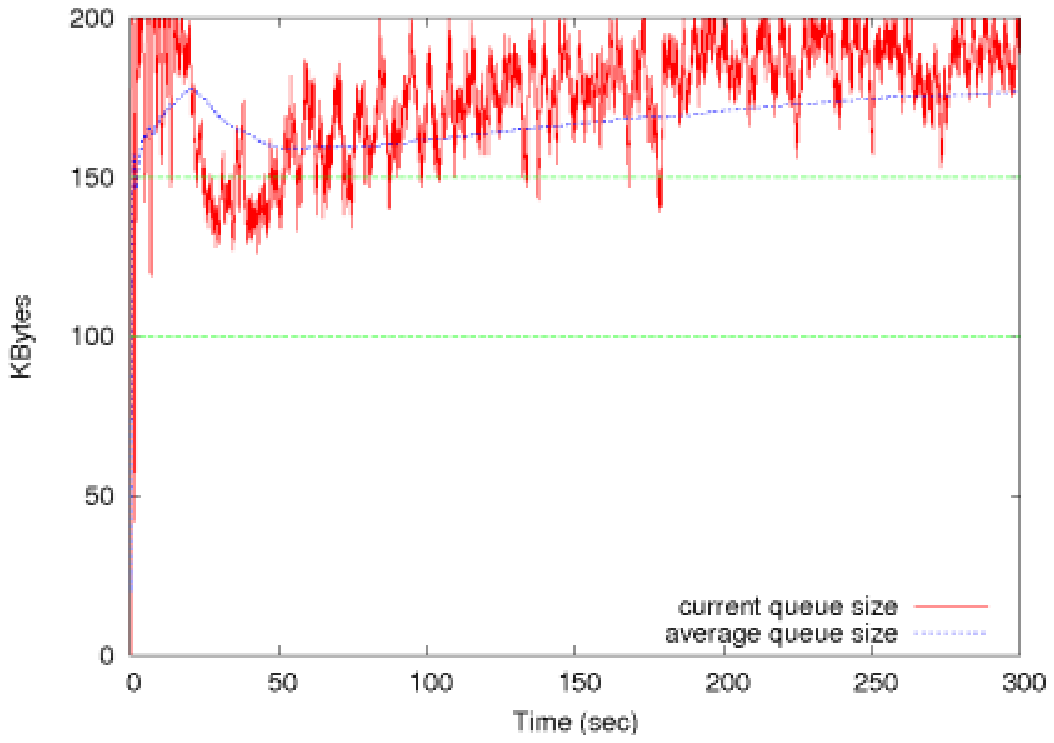}
\label{fig:red1}
}
\subfigure[FRED]{
\includegraphics[ keepaspectratio=true, width=\imgsz, angle =-90]{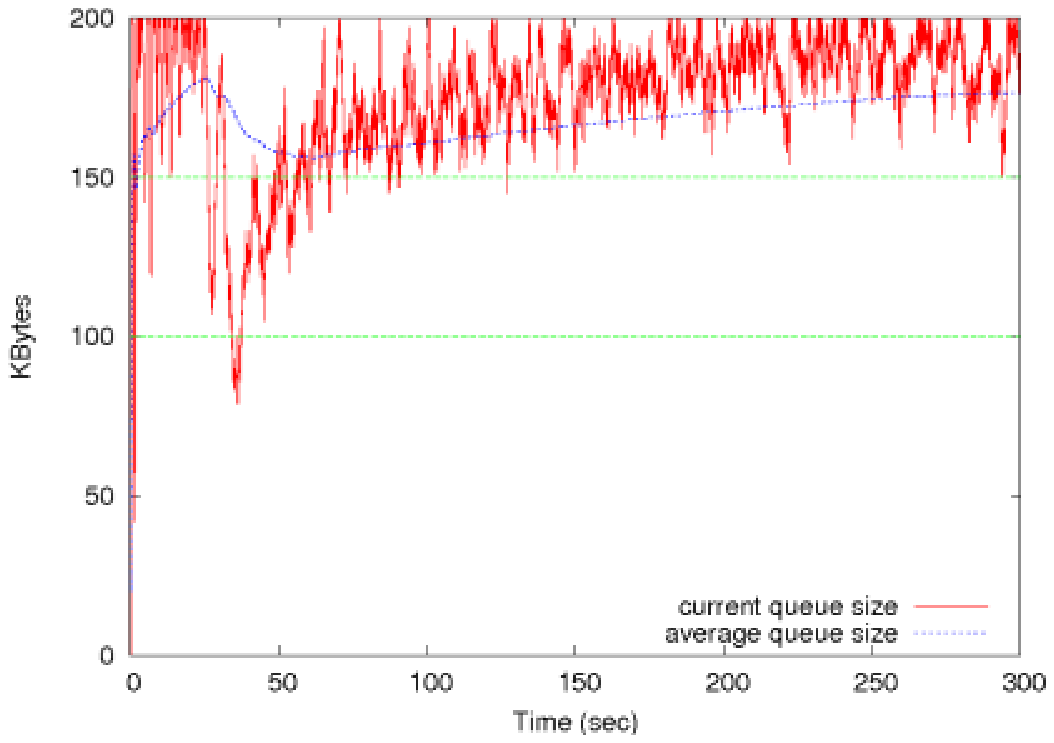}
\label{fig:fred1}
}
\subfigure[ARED]{
\includegraphics[ keepaspectratio=true, width=\imgsz, angle =-90]{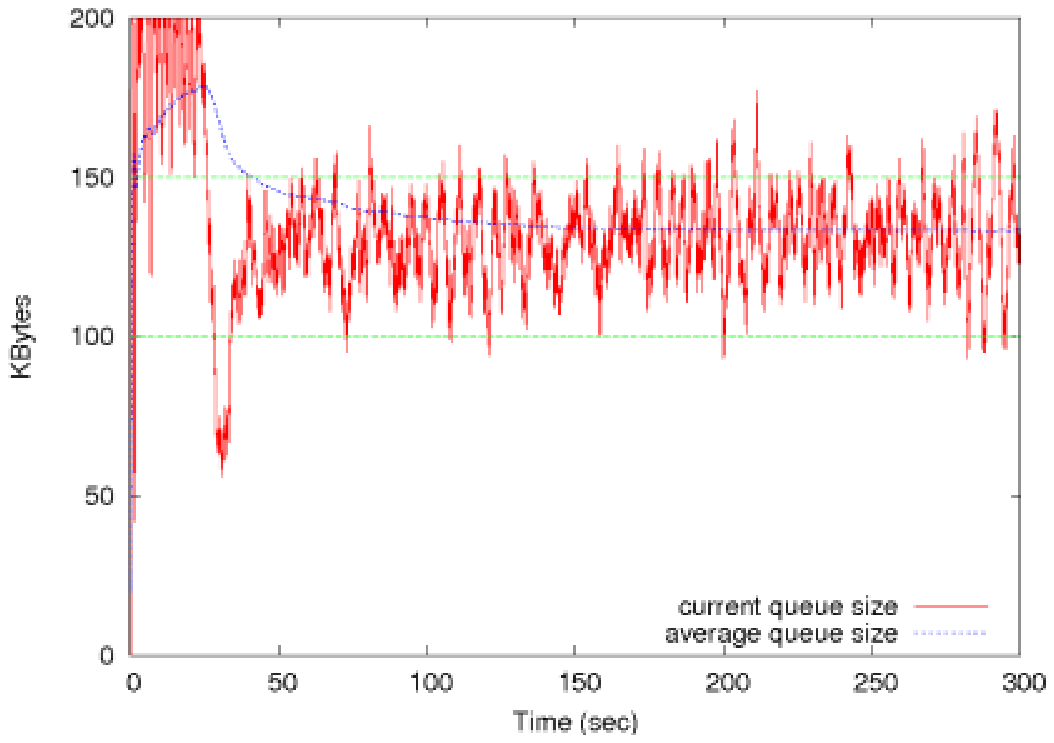}
\label{fig:ared1}
}
~\\
\subfigure[PI]{
\includegraphics[ keepaspectratio=true, width=\imgsz, angle =-90]{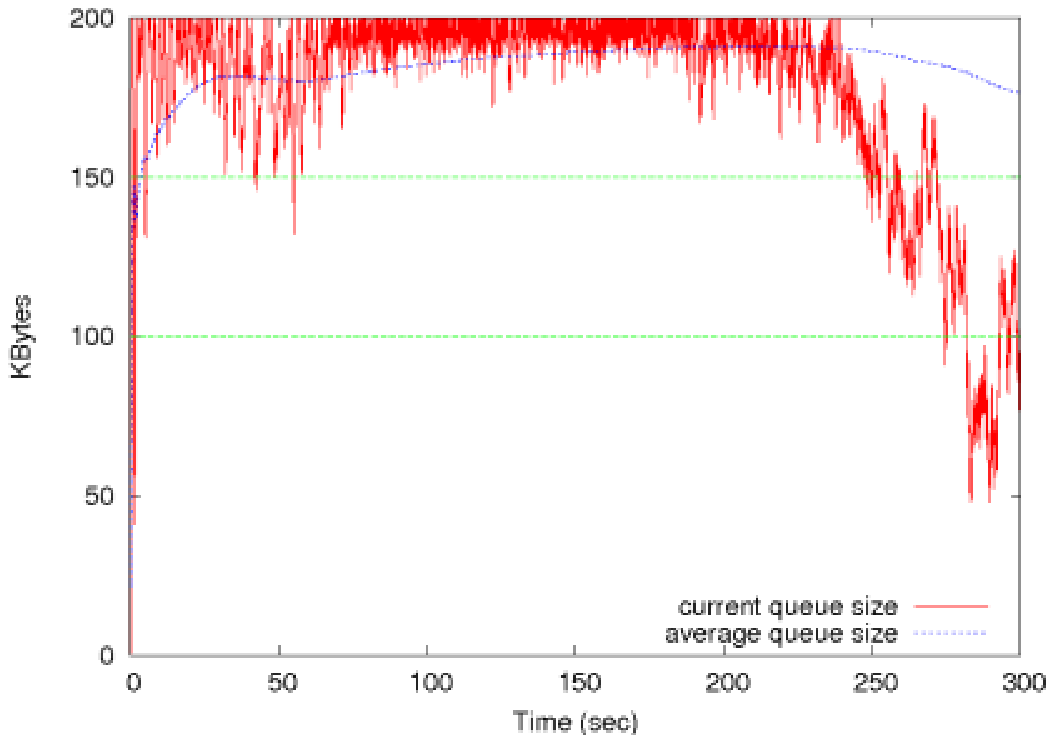}
\label{fig:pi1}
}
\subfigure[KRED]{
\includegraphics[ keepaspectratio=true, width=\imgsz, angle =-90]{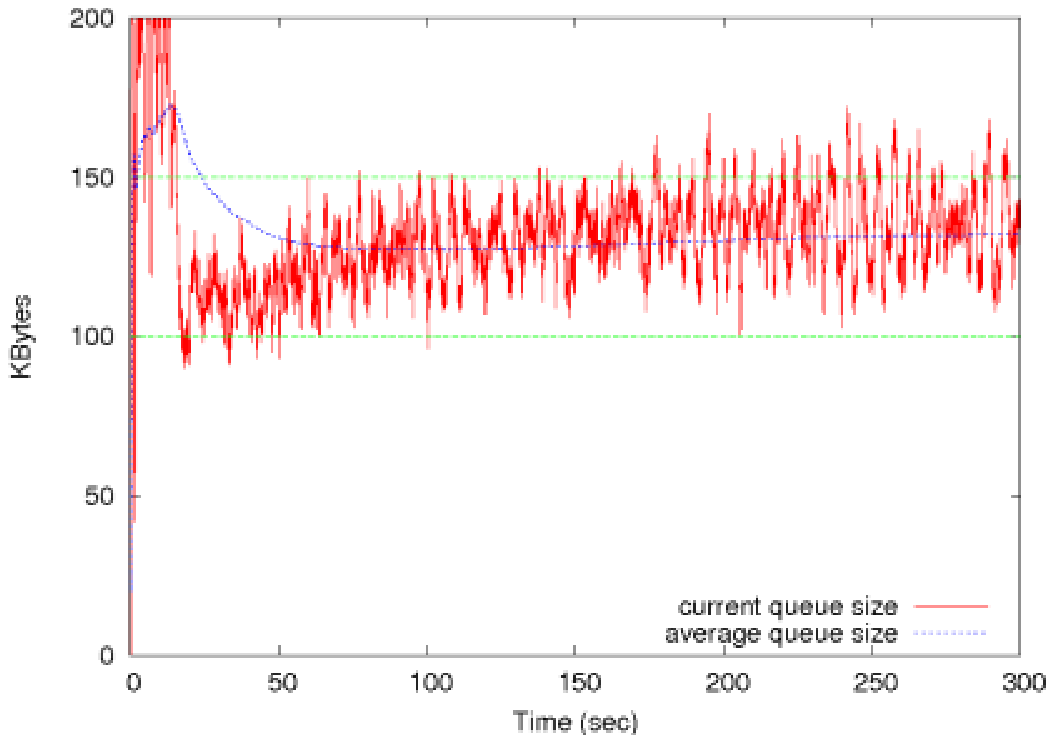}
\label{fig:kred1}
}
\caption{Performance comparison of various AQM with KRED, (1\up{st} scenario)}
\label{fig:tests1}
\end{center}
\end{figure*} 

Results are given in figure \ref{fig:tests1}. Each graph shows the instantaneous and average queue size. 
The two horizontal lines represent the $min_{th}$ and $max_{th}$ threshold RED parameters. 
The results presented for the KRED queue are obtained after the training process.
As shown in these figures, ARED \ref{fig:ared1} and KRED \ref{fig:kred1} queues obtain a stable queue length between both thresholds compared to the
others. 
To better illustrate the benefit of our algorithm, we use as comparative metric the queue delay. The link statistics reported table \ref{tab:stats1} 
show that the queueing delay for ARED and KRED are the lowest. The drop rate for both
queue is similar and the overall throughput at the output link is equal for all AQM.

With this first experiment we can conclude that ARED and KRED are the best algorithms in terms of stability of the queue when 
the traffic load increases but we cannot stand in favor of KRED since the results obtained are in the same order of magnitude. Indeed, 
the overall performances obtained by both AQM are quite similar. However, we have to keep in mind that fixed initial parameters are needed 
for ARED (given in table \ref{tab:params}).

The second scenario, presented in the next section, extends these measurements by changing the traffic pattern during the
simulation and the RTT of each flow is ranging from 64 to 102 ms. The original parameters remain unchanged in order to verify the well adaptability of these AQM to the rapid traffic change
problem.

\begin{table}
\begin{center}
\begin{tabular}{|c||c|c|c|}
\hline
AQM & Mean / Std. Dev. & Mean / Std. Dev.    & TCP       \\
    & Queue Delay (ms) & Link Throughput (Mbit/s) & drop rate \\\hline
RED  & 29.17 / 3.05 & 4.9978 / 0.29 & 22.23\%\\
FRED & 29.09 / 3.55 & 4.9978 / 0.29 & 22.39\%\\
ARED & 22.24 / 3.65 & 4.9978 / 0.29 & 25.23\%\\
PI   & 29.51 / 5.27 & 4.9979 / 0.29 & 22.53\%\\
KRED & 21.92 / 3.02 & 4.9978 / 0.29 & 25.34\%\\\hline
\end{tabular}
\end{center}
\caption{Statistics from 1\up{st} scenario}
\label{tab:stats1}
\end{table}

\subsection{Second scenario}

\begin{figure*}[ht!]
\begin{center}
\subfigure[RED]{
\includegraphics[ keepaspectratio=true, width=\imgsz, angle =-90]{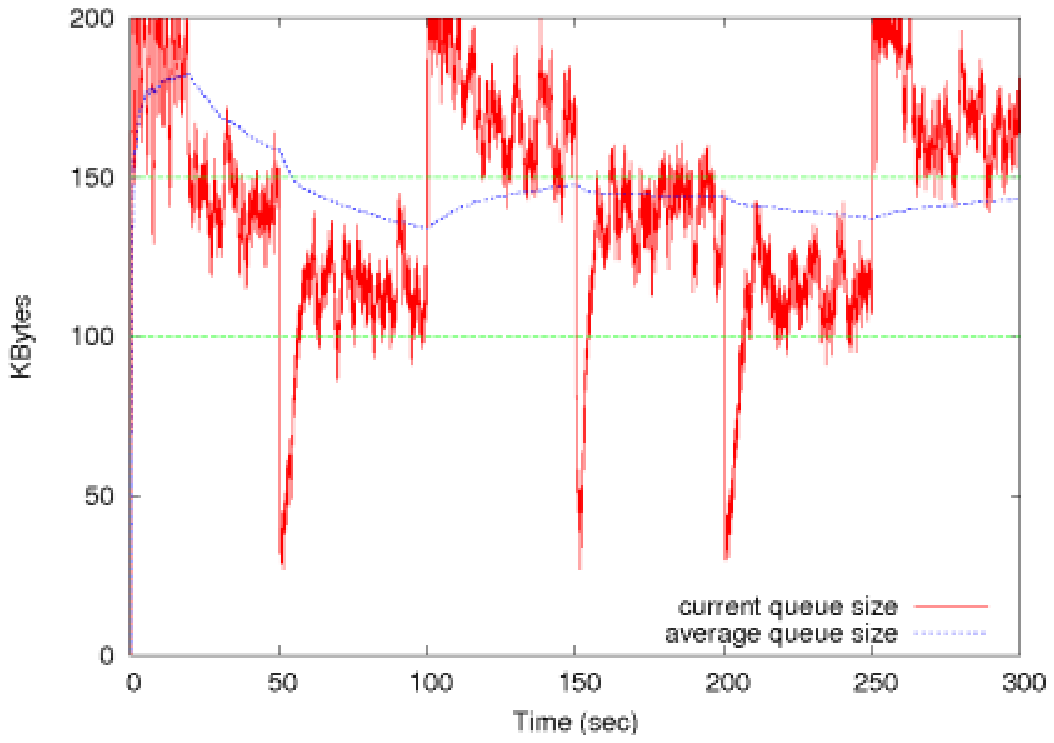}
\label{fig:red2}
}
\subfigure[FRED]{
\includegraphics[ keepaspectratio=true, width=\imgsz, angle =-90]{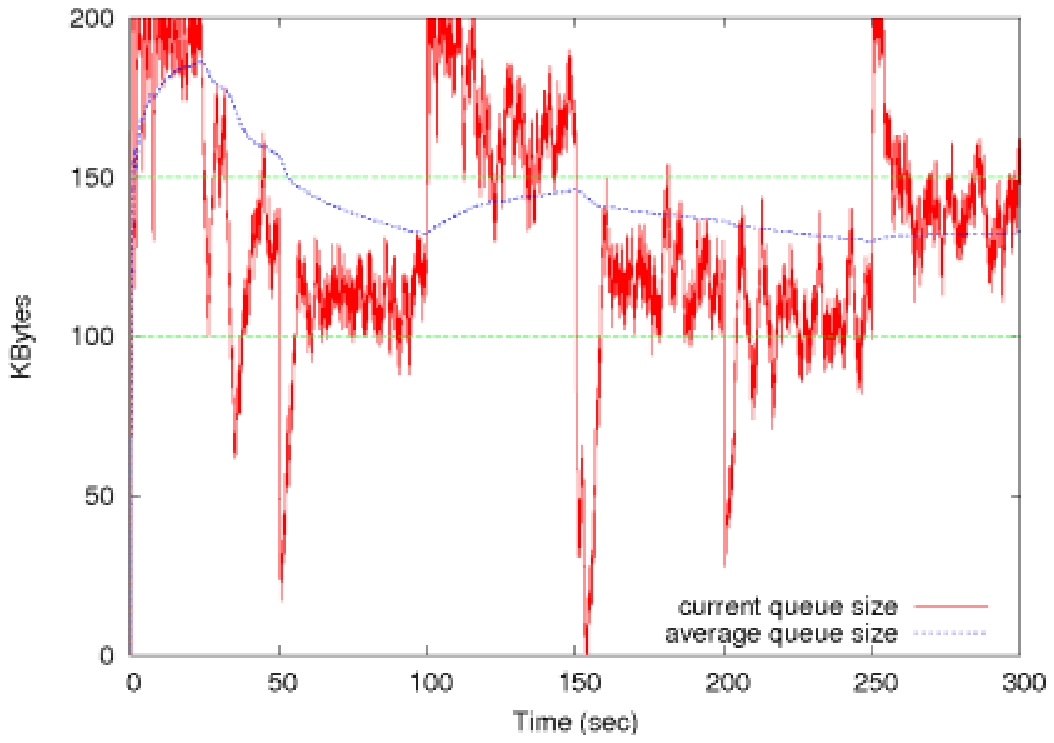}
\label{fig:fred2}
}
\subfigure[ARED]{
\includegraphics[ keepaspectratio=true, width=\imgsz, angle =-90]{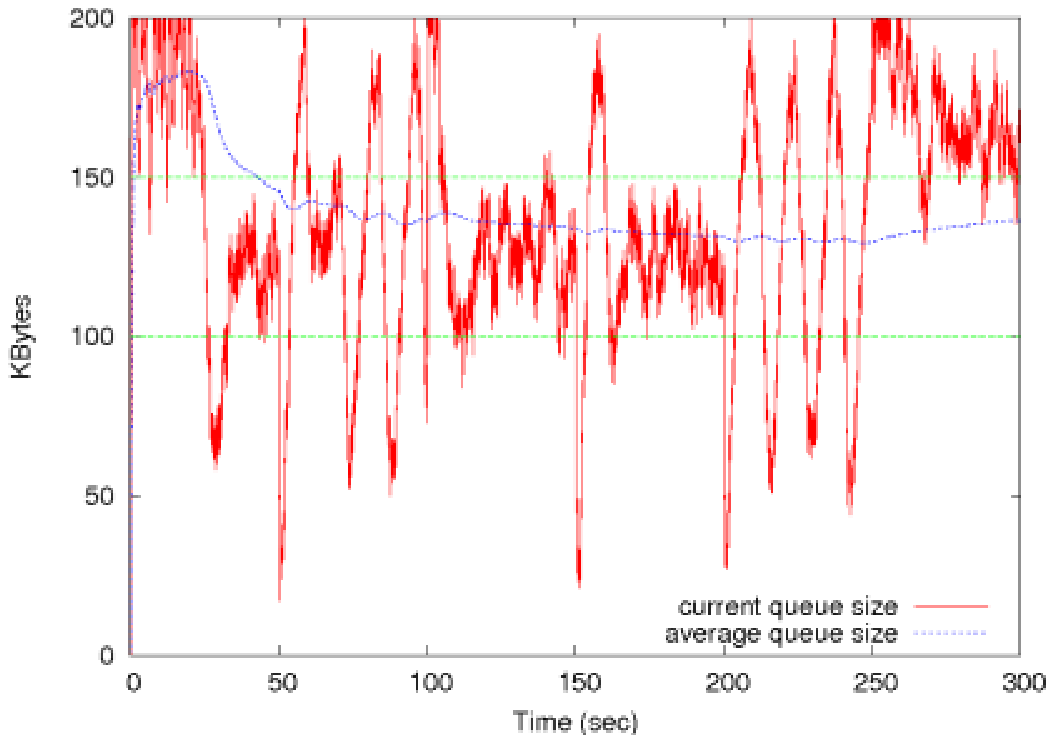}
\label{fig:ared2}
}
~\\
\subfigure[PI]{
\includegraphics[ keepaspectratio=true, width=\imgsz, angle =-90]{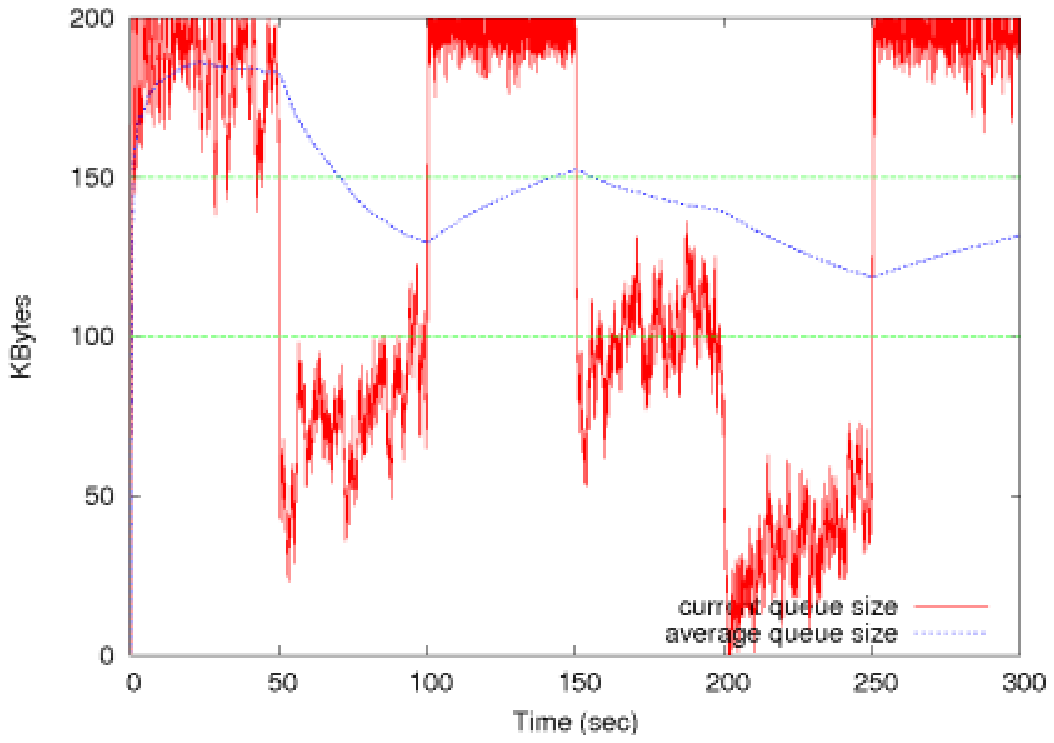}
\label{fig:pi2}
}
\subfigure[KRED]{
\includegraphics[ keepaspectratio=true, width=\imgsz, angle =-90]{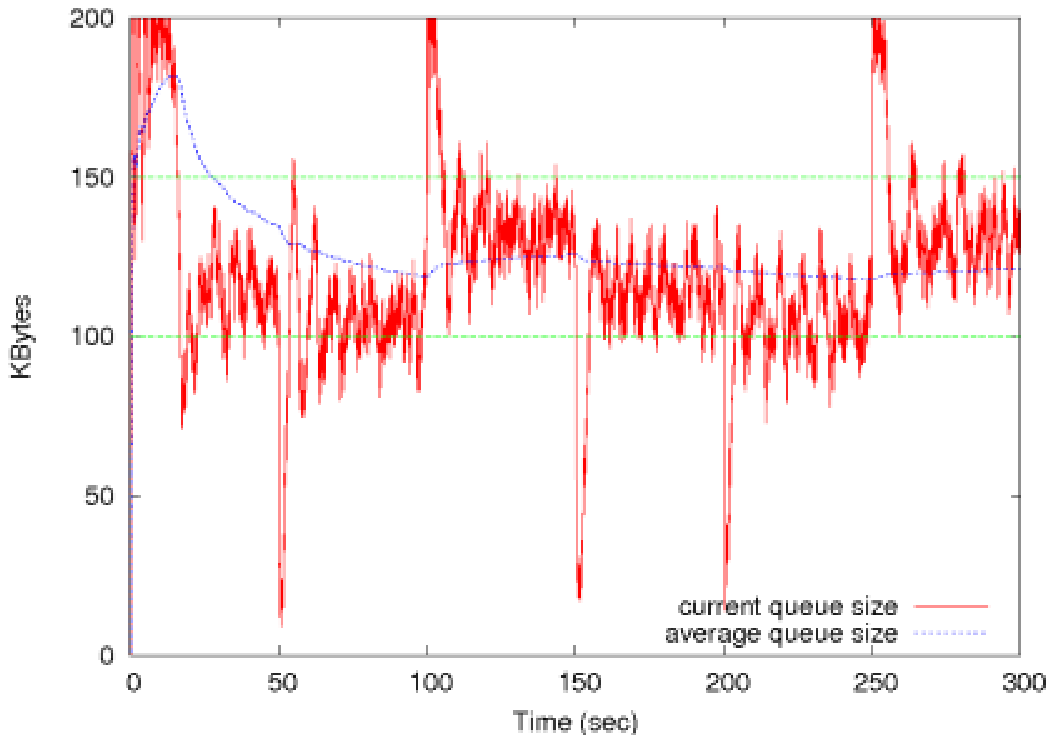}
\label{fig:kred2}
}
\caption{Performance comparison of various AQM with KRED (2\up{nd} scenario)}
\label{fig:tests2}
\end{center}
\end{figure*} 

As shown in figure \ref{fig:tests2}, the KRED queue obtains a stable queue length between both thresholds compared to the
AIMD process method of the FRED \ref{fig:fred2} and ARED \ref{fig:ared2}. Moreover, KRED reacts rapidly to a traffic change
compared to ARED.
Due to the rapid traffic changing, the $max_{p}$ value is constantly recomputed and the previous computed value
strongly impacts on the current result. In the case of an AIMD process to compute the best $max_{p}$ value, if the weights 
are small, the pace of convergence to the optimal value is slow and if the weights are high, the resulting probability can strongly 
oscillate when the traffic is changing. The initial configuration parameters used with success in the first scenario by ARED are not
adapted to the second one.
Thus, the KRED proposal allows to overcome this difficult problem of initial setting which is managed by the neural network.
Finally, table \ref{tab:stats2} give the statistics of this scenario and show that KRED still obtains
the lowest average queuing delay.

\begin{table}
\begin{center}
\begin{tabular}{|c||c|c|c|}
\hline
AQM & Mean / Std. Dev. & Mean / Std. Dev.    & TCP       \\
    & Queue Delay (ms) & Link Throughput (Mbit/s) & drop rate \\\hline
RED  & 23.68 / 5.68  & 4.9978 / 0.0294 & 7.65\%\\
FRED & 21.96 / 6.14  & 4.9978 / 0.0294 & 8.54\%\\
ARED & 22.73 / 6.27  & 4.9978 / 0.0294 & 8.42\%\\
PI   & 21.67 / 10.78 & 4.9979 / 0.0294 & 8.70\%\\
KRED & 20.10 / 4.75  & 4.9978 / 0.0294 & 9.15\%\\\hline
\end{tabular}
\end{center}
\caption{Statistics from 2\up{nd} scenario}
\label{tab:stats2}
\end{table}

\section{Discussion and conclusions}
\label{sec:conclusion}

This paper introduces Kohonen-RED: an adaptive RED mechanism easily implementable. The idea deals with the use of a Kohonen neural
network to compute the optimal probability parameter in order to achieve stable queue length. KRED reduces the number of parameters and
in particular the non obvious ones. 
The Kohonen network does not need to be retrained and therefore can be put in hardware in the context of a router implementation.
The mechanism's efficiency has been illustrated through ns-2 simulation where other schemes fail. In this work, we use 
a Kohonen based neural network specifically designed to solve the pole balancing problem. One of the main contribution of this study is to
show the feasibility of using neural network to solve a networking stability problem.
Considering promising preliminary results, we are currently designing a specific neural network for RED queue able to stabilize on a given
value and not between two bounds and we are investigating on the improvements required in the core mechanism itself to achieve this goal.
We also explore the design of a neural network able to accurately characterize the TCP behaviour.

\section*{Acknowledgments}

The authors would like to thank Sebastien Ardon and Guillaume Jourjon and Max Ott for the discussion about this mechanism
and the support of the National ICT Australia (NICTA). 

%------------------------------------------------------------------------- 
\bibliographystyle{plain}
\bibliography{biblio}
\end{document}

%% file: img/red-pole.pstex_t
\begin{picture}(0,0)%
\includegraphics{./img/red-pole.pstex}%
\end{picture}%
\setlength{\unitlength}{3947sp}%
\begingroup\makeatletter\ifx\SetFigFontNFSS\undefined%
\gdef\SetFigFontNFSS#1#2#3#4#5{%
  \reset@font\fontsize{#1}{#2pt}%
  \fontfamily{#3}\fontseries{#4}\fontshape{#5}%
  \selectfont}%
\fi\endgroup%
\begin{picture}(9247,2909)(1336,-4726)
\put(3526,-4561){\makebox(0,0)[lb]{\smash{{\SetFigFontNFSS{17}{20.4}{\sfdefault}{\mddefault}{\updefault}{\color[rgb]{0,0,0}$max_{th}$}%
}}}}
\put(5551,-4561){\makebox(0,0)[lb]{\smash{{\SetFigFontNFSS{17}{20.4}{\sfdefault}{\mddefault}{\updefault}{\color[rgb]{0,0,0}$min_{th}$}%
}}}}
\put(6376,-3136){\makebox(0,0)[lb]{\smash{{\SetFigFontNFSS{17}{20.4}{\sfdefault}{\mddefault}{\updefault}{\color[rgb]{0,0,0}$max_p$}%
}}}}
\end{picture}%